\documentclass{svproc}
\usepackage{url}

\usepackage{graphicx}
\usepackage{multicol}
\usepackage{amsmath,amssymb,amsfonts}
\usepackage[symbol]{footmisc}
\usepackage{xcolor}%

\begin{document}
\title{Exploring Layered Structure Inside Earth Using Atmospheric Neutrino Oscillation at IceCube DeepCore}
\titlerunning{Exploring layered structure inside Earth}
%
\author{J Krishnamoorthi\inst{}\footnote[2]{also at Institute of Physics, Sachivalaya Marg, Sainik School Post, Bhubaneswar 751005, India and Dept. of Physics, Aligarh Muslim University, Aligarh 202002, India.}  \\
(For the IceCube Collaboration\footnote[1]{\url{http://icecube.wisc.edu}})}
\authorrunning{J Krishnamoorthi (For the IceCube Collaboration)} 
%
\tocauthor{For the IceCube Collaboration}
\institute{Dept. of Physics and Wisconsin IceCube Particle Astrophysics Center, University of Wisconsin-Madison, Madison, WI 53706, USA \\
\email{kjayakumar@icecube.wisc.edu}}

\maketitle              
\begin{abstract}
The IceCube detector, using its densely instrumented center, called DeepCore, can detect multi-GeV atmospheric neutrinos. The oscillation pattern of neutrinos is altered due to interactions with ambient electrons as they pass through Earth. The changes in these patterns are influenced by the amount of matter and its specific arrangement. As neutrinos propagate, they retain information about the densities they encounter. Our study demonstrates that IceCube DeepCore can utilize the Earth's matter effects to distinguish between a homogeneous matter density profile and a layered structure density profile of Earth. In this contribution, we present that IceCube DeepCore data equivalent to 9.3 years of observation can reject the homogeneous matter density profile with a confidence level of 1.4$\sigma$.

\end{abstract}

\section{Introduction}

The information about the interior of Earth primarily comes from the gravitational and seismic studies. These measurements involve indirect probes due to the extreme temperature and pressure conditions inside Earth. The gravitationally measured values of mass and moment of inertia of Earth indicate that the density distribution inside Earth is not uniform. Rather, Earth has a higher density at its center. On the other hand, using the seismic wave velocity data, the widely used density model of Earth, the Preliminary Reference Earth Model (PREM)~\cite{Dziewonski:1981xy}, has been developed. 

The cosmic rays entering Earth's atmosphere interact with air nuclei, producing a cascade of charged particles. The decay of these particles results in the generation of atmospheric neutrinos, primarily the electron and muon flavors. These atmospheric neutrinos cover a wide range of energies, from a few MeV to more than a TeV, and a wide range of baselines. During the propagation, the neutrino changes its flavor from the initially produced one; this phenomenon is known as neutrino oscillation. As multi-GeV atmospheric neutrinos travel through Earth, they experience an additional potential that alters their oscillation probabilities due to coherent forward scattering with ambient electrons. This modification depends on the amount and distribution of matter encountered during their propagation inside Earth, and hence, it can be used to probe the interior of Earth. This method is known as neutrino oscillation tomography.

At high energies, the neutrino-nucleon cross section increases, leading to neutrino absorption inside Earth. Studying these neutrinos could also provide complementary information about the inner structure of Earth, and this method is known as neutrino absorption tomography. Using the neutrino absorption inside Earth, the authors in Ref.~\cite{Gonzalez-Garcia:2007wfs} estimated the IceCube detector's ability to reject Earth's homogeneity with a confidence of 3.4$\sigma$. 

In this work, we focus only on neutrino oscillation tomography. Our study aims to establish that Earth has a layered density structure by rejecting the homogeneous matter density hypothesis with respect to 12-layered PREM profile using the multi-GeV atmospheric neutrino oscillation data observed at IceCube DeepCore.

\section{IceCube DeepCore Detector and Event Sample}
The IceCube Neutrino Observatory~\cite{IceCube:2016zyt} is a Cherenkov-based neutrino detector consisting of 5,160 Digital Optical Modules (DOMs) deployed in the South Pole ice. The bottom central part of the IceCube detector is known as DeepCore, a sub-array designed to improve the detection ability in multi-GeV energy range. 
The DOMs can detect the Cherenkov photons emitted by the secondary charged particle produced in the interactions of neutrinos with ice. The DOMs provide the number of photons and their arrival time, which is used later for filtering and reconstruction. Multiple filter levels are used to remove the background contamination from random detector noise and the atmospheric muons. Once the amount of background is reduced significantly, we apply separate Convolutional Neural Networks (CNNs) to reconstruct the energy and the arrival direction of neutrinos, to identify particle types (PID), and to remove atmospheric muon background. The CNN PID classifier is used to look for a $\nu_{\mu}$ charged-current interactions signature by exploiting the minimum-ionizing property of muon, which leaves a signal in an elongated pattern across the detector.

We use the IceCube DeepCore data~\cite{IceCube:2024xjj} equivalent to 9.3 years for this analysis. 
In this data sample, the reconstructed energy ranges from 3 to 100 GeV, and the reconstructed $\cos(\theta_{zenith})$ ranges from -1 to 0, corresponding to the upward-going neutrinos. Based on the PID score, the sample is classified into cascade-like, mixed, and track-like bins with the boundary [0.0, 0.33, 0.39, 1.0].

\section{Analysis}
In this section, we discuss the analysis methodology. The analysis is first performed with the simulated data of IceCube DeepCore, which is equivalent to 9.3 years of observations, to check its robustness. The reconstructed events are binned as a function of reconstructed energy, $\cos(\theta_{zenith})$, and PID. The simulated Monte Carlo (MC) events are re-weighted using the oscillation probability for the different Earth density models (12-layered PREM or uniform density profile). The Asimov and median sensitivities are calculated based on the definition from Ref.~\cite{ciuffoliSensitivityNeutrinoMass2014a} and presented as a function of simulated $\sin^2\theta_{23}$, since the sensitivity depends on the large uncertainty on $\theta_{23}$. In the fit, the oscillation parameters, $\theta_{23}$, and $\Delta m_{31}^2$ are kept free, while the other oscillation parameters; $\theta_{12}$, $\theta_{13}$, and $\Delta m_{21}^2$ are kept fixed. Since the sensitivity does not depend on the choice of $\delta_{\text{CP}}$, we kept it fixed at zero. Apart from the oscillation parameters, we also consider various systematic parameters, which include uncertainty in flux, cross section, detector effect, and atmospheric muon background~\cite{IceCubeCollaboration:2023wtb}. 

\begin{figure}
    \centering
    \includegraphics[width=0.485\linewidth]{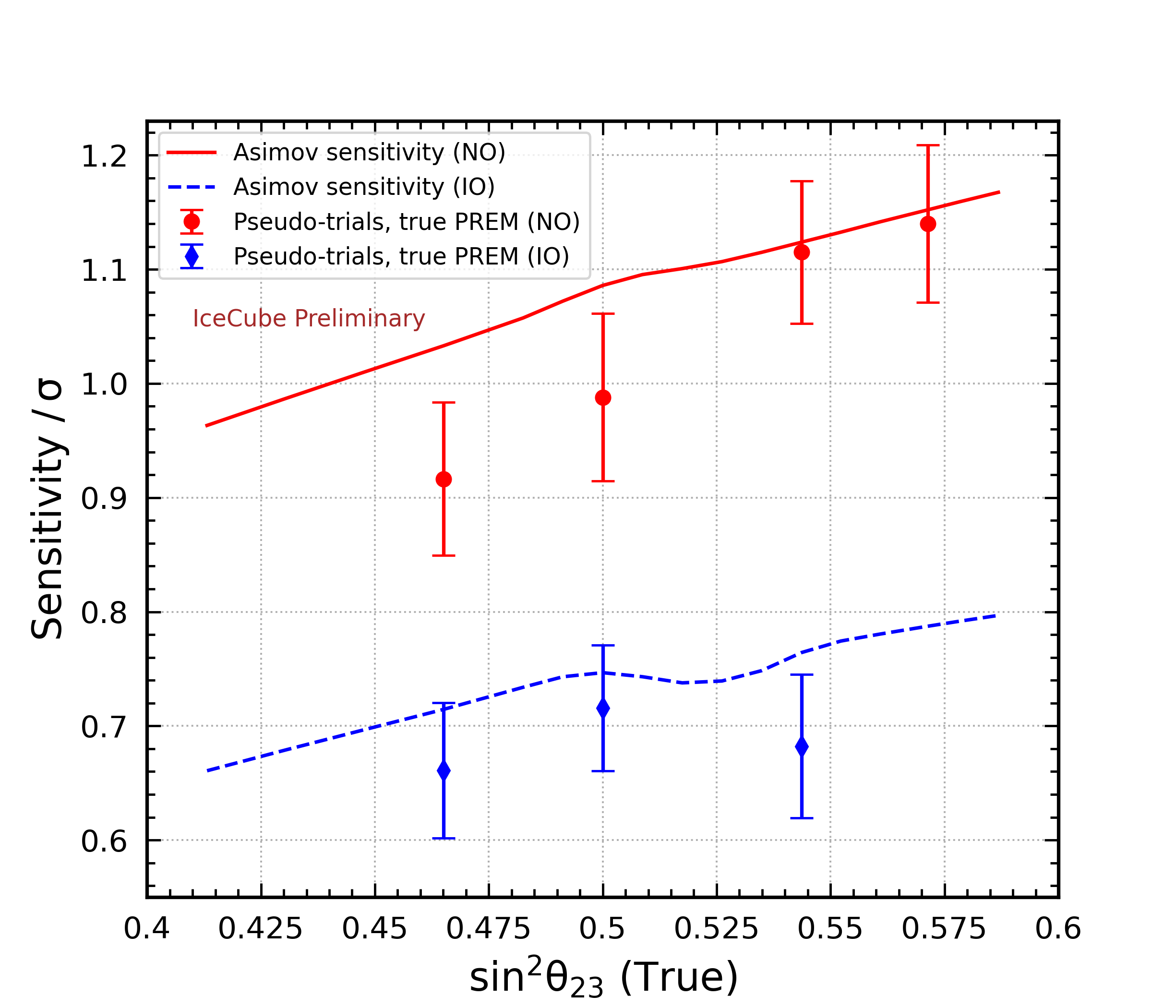}
    \includegraphics[width=0.485\linewidth]{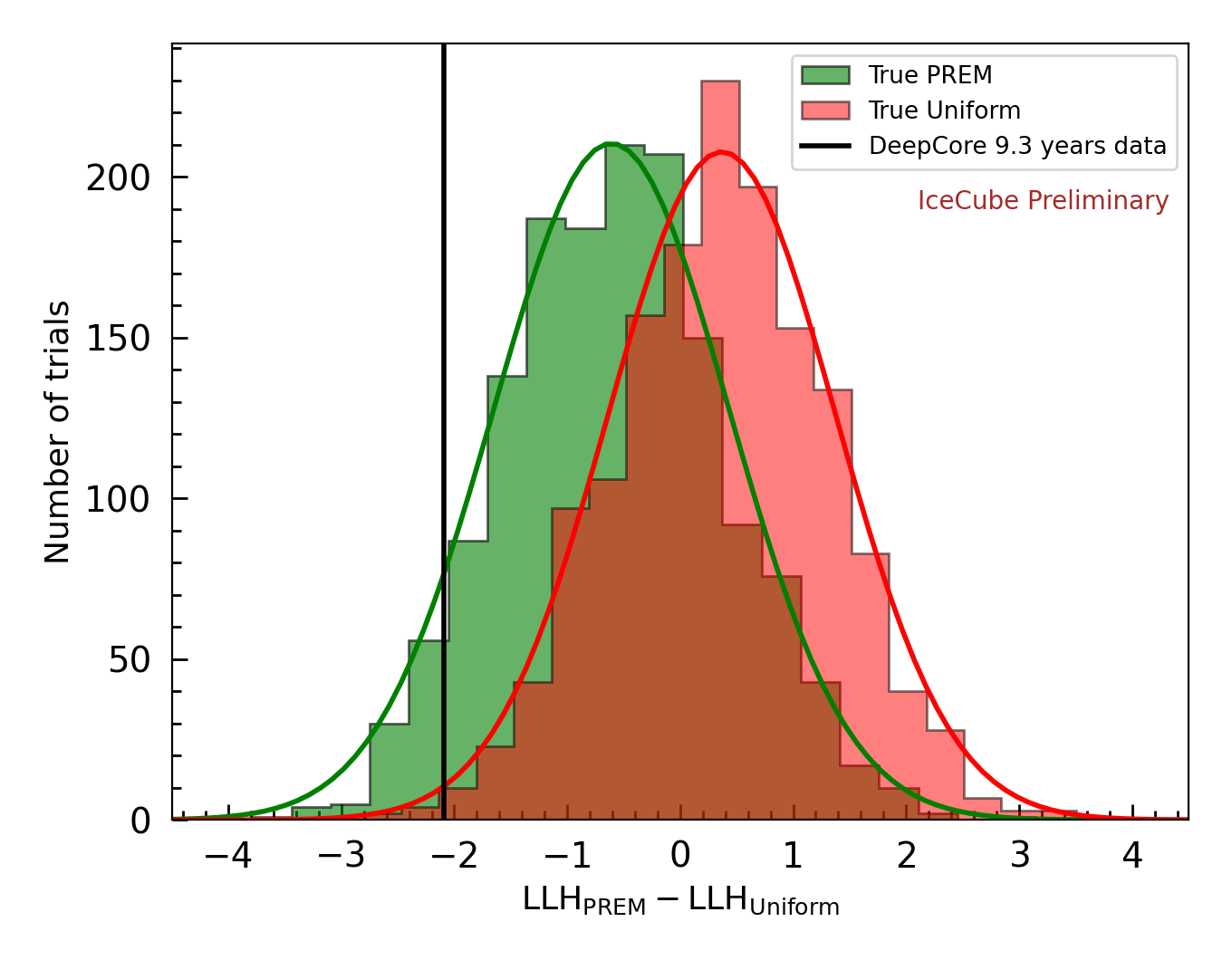}
    \caption{Left: Sensitivity to reject Uniform hypothesis as a function of true choices of $\sin^2\theta_{23}$ calculated using Asimov-method (curves), and median sensitivity calculated using the frequentist method at few choices of $\sin^2\theta_{23}$ (markers). The error bars on the frequentist points represent the standard error. The red (blue) color corresponds to the assumption of normal (inverted) neutrino mass ordering. Right: Distributions of metric (true PREM and true Uniform) are calculated from the pseudo-experiments, which are generated with the best-fit parameters from the hypotheses PREM (green) and Uniform (red). The vertical black line represents the observed experimental value of the metric. }
    \label{sensitivity_plot}
\end{figure}

The left panel of Fig. \ref{sensitivity_plot} shows the sensitivity as a function of simulated $\sin^2\theta_{23}$ of IceCube DeepCore to reject the uniform hypothesis with respect to a 12-layered PREM hypothesis. The sensitivity shows an increasing trend because of the $P(\nu_\mu \rightarrow \nu_\mu)$ survival probability and the $P(\nu_\mu \rightarrow \nu_e)$ appearance probability being proportional to $\sin^2\theta_{23}$, as shown by a series expansion in Ref.~\cite{Akhmedov:2004ny} and is also dependent on the choice of neutrino mass ordering. The Asimov sensitivity shows better agreement with the median sensitivity, which is calculated for a few choices of $\sin^2\theta_{23}$. The experimental method of finding the preferred hypothesis by the data and the significance of rejecting the opposite hypothesis is followed from Ref.~\cite{aartsenDevelopmentAnalysisProbe2020}.

\section{Results and Conclusion}
In this section, we discuss the experimental results of IceCube DeepCore to reject the homogeneous Earth density model. In the right panel of Fig. \ref{sensitivity_plot} shows the observed $\Delta LLH = LLH_\text{PREM} - LLH_\text{Uniform}$ value and the distributions obtained from the statistically fluctuated MC simulation with the assumption of 12-layered PREM and uniform density model. The $LLH_\text{PREM}$ and $LLH_\text{Uniform}$ represent the log-likelihood value obtained by fitting the experimental data (simulated) with the PREM and Uniform hypotheses. The observed p-values for PREM and Uniform hypothesis are around 94\% and 0.46\%, respectively. Following the CL$_s$ definition from Ref.~\cite{Qian:2014nha}, the confidence level to reject the Uniform hypothesis is 92.4\%. The significance of rejecting the Uniform hypothesis with respect to the 12-layered PREM profile is 1.4$\sigma$. This result demonstrates the methodology that atmospheric neutrino oscillation could be a potential candidate for studying the internal properties of Earth. The upcoming IceCube Upgrade~\cite{IceCube:2023ins} will enhance the sensitivity of this study by utilizing the low energy threshold and better detector calibration.

\subsubsection{\ackname}  We acknowledge the financial support from the Dept. of Atomic Energy (DAE), Govt. of India and the Swarnajayanti Fellowship provided by the Science and Engineering Research Board (SERB), Govt. of India.

%
%

%
\bibliographystyle{spphys}
\bibliography{References}
\end{document}